\documentclass[a4paper,11pt]{article}
\pdfoutput=1

\usepackage{jheppub,amsfonts,mathrsfs,caption,subcaption}

\usepackage{amsmath}
\usepackage{amssymb}
\usepackage{hyperref}
\usepackage{enumitem}
\usepackage{color}
\usepackage{cleveref}
\crefname{equation}{equation}{equations}
\Crefname{equation}{Equation}{Equations}
\crefrangelabelformat{equation}{(#3#1#4--#5#2#6)}

\crefmultiformat{equation}{(#2#1#3}{, #2#1#3)}{#2#1#3}{#2#1#3}
\Crefmultiformat{equation}{(#2#1#3}{, #2#1#3)}{#2#1#3}{#2#1#3}

\numberwithin{equation}{section}

\title{Holographic relations for OPE blocks in excited states}

\author[a]{Jesse C. Cresswell}
\author[a]{Ian T. Jardine}
\author[a,b]{Amanda W. Peet}

\affiliation[a]{Department of Physics, University of Toronto, \\ 60 St. George St., Toronto, Canada}
\affiliation[b]{Department of Mathematics, University of Toronto,\\ 40 St. George St., Toronto, Canada}

\emailAdd{jcresswe@physics.utoronto.ca}
\emailAdd{itjardine@hotmail.com}
\emailAdd{awpeet@physics.utoronto.ca}

\abstract{We study the holographic duality between boundary OPE blocks and geodesic integrated bulk fields in quotients of AdS$_3$ dual to excited CFT states. The quotient geometries exhibit non-minimal geodesics between pairs of spacelike separated boundary points which modify the OPE block duality. We decompose OPE blocks into quotient invariant operators and propose a duality with bulk fields integrated over individual geodesics, minimal or non-minimal. We provide evidence for this relationship by studying the monodromy of asymptotic maps that implement the quotients.}


\arxivnumber{1809.09107}

\begin{document}
\maketitle
\flushbottom


\section{Introduction}

Since its initial formulation, the AdS/CFT correspondence has opened up many new avenues for studying gravity \cite{Maldacena1999}. It provides a dictionary that can translate unfamiliar gravitational physics into familiar field theory, and vice versa. One of its most powerful aspects is the ability to encode the spatial organization of the bulk as a relationship between the degrees of freedom in the CFT. A particularly useful way of analyzing the geometry of spacetime is through examining the structure of geodesics and extremal surfaces. This has a long history 
in the AdS/CFT context, and an important new theme was begun with the work of \cite{Ryu2006}. Their results in AdS${}_3$ showed that the entanglement entropy of a CFT$_2$ interval is dual to the length of a bulk geodesic anchored at the interval's endpoints. 

The connection between entanglement and geometry \cite{VanRaamsdonk2010} has become of fundamental interest, and has been expanded to many other aspects of quantum information. These include the emergence of gravitational equations of motion from CFT entanglement entropies \cite{Faulkner2017}, bulk gauge freedom interpreted as boundary quantum error correcting codes \cite{Almheiri2015,Mintun2015,Pastawski2015}, the volume of Einstein-Rosen bridges as complexity \cite{Susskind2014}, and the entanglement wedge cross section as CFT entanglement of purification \cite{Umemoto2018}. 

A useful auxiliary space termed kinematic space has been introduced describing the structure of geodesics while also geometrizing entanglement entropy \cite{Czech2015,Czech2016,Boer2016}. Each boundary anchored geodesic, or equivalently each pair of boundary points, is viewed as a single point in kinematic space. One of the major developments discovered through this construction was the holographic dual of a bulk field integrated over a boundary anchored geodesic, namely the OPE block of the corresponding dual operator in the CFT. This is closely related to the duality between conformal blocks in the CFT and geodesic Witten diagrams in the bulk \cite{Hijano2016,Hijano2015}. The properties of OPE blocks themselves have been studied further for defect CFTs \cite{Fukuda2018, Kobayashi2018} and using modular flow \cite{Sarosi2018}.

While these works on kinematic space were thorough, they mainly focused on pure AdS. Followup papers \cite{Czech2016a,Asplund2016,Zhang2017,Karch2017,Cresswell2017,Abt2017,Abt2018} have worked towards extending kinematic space and the OPE block duality to more general AdS spacetimes. We will continue this line of inquiry for AdS${}_3$, where all vacuum solutions to the Einstein equation with negative cosmological constant are locally AdS${}_3$ and can be obtained as quotients. The immediate challenge is that there is no longer a unique geodesic through the bulk between any pair of boundary endpoints. A natural question is to ask how the CFT dual of a geodesic integrated bulk field changes. We will argue that in states dual to quotient geometries, OPE blocks decompose into contributions which are invariant under the quotient action. Each contribution is dual to a bulk field integrated over a single geodesic which may wind around the quotient's fixed points. 

Our arguments are based on the monodromy of maps between pure AdS$_3$ and the quotient geometries. In the bulk the monodromy is responsible for the appearance of non-minimal geodesics, and on the boundary it induces non-analyticities in the OPE blocks. We resolve the latter issue by constructing quotient invariant OPE blocks, and interpret them in terms of winding geodesics. We often utilize the group manifold description of AdS$_3$ and its quotients, in which the structure of geodesics is made clear, and their lengths are easily computable. Throughout, we work with the Euclidean and Lorentzian versions of the construction in parallel to emphasize their differences.

In Section \ref{sec:2} we review the duality between OPE blocks and geodesic integrated bulk fields. Then we introduce the quotient spacetimes of interest and find explicit maps between them and pure AdS$_3$. In Section \ref{sec:3} we use these maps to study the structure of geodesics in the quotient geometries and determine their lengths. In Section \ref{sec:4} we construct quotient invariant OPE blocks, highlighting their relationship to winding geodesics. In Section \ref{sec:5} we conclude with a summary and discussion of remaining open questions.
 
\section{Preliminaries}\label{sec:2}
\subsection{OPE blocks and kinematic space}

In a 2d CFT, the OPE allows us to expand the product of two quasiprimary operators in terms of a basis of local operators at a single location. The OPE can be organized by the contributions from conformal families in the theory, each consisting of a quasiprimary $\mathcal{O}_k$ and its descendants. Considering two scalar operators with the same conformal weight $\Delta$, conformal symmetry dictates that
\begin{equation}\
\mathcal{O}_{i}(x)\mathcal{O}_{j}(0) =  \sum_{k}C_{ijk}\left|x\right|^{\Delta_{k}-2\Delta}\big(1+b_{1}\,x^{\mu}\partial_{\mu}+b_{2}\,x^{\mu}x^{\nu}\partial_{\mu}\partial_{\nu}+\ldots\big)\mathcal{O}_{k}(0)\,,
\end{equation}
with some theory dependent constants $C_{ijk}$, and theory independent constants $b_i$. Since much of this structure is fixed by symmetry, it is convenient to define an OPE block ${\mathcal{B}}^{ij}_k(x_i,x_j)$ associated to each quasiprimary $\mathcal{O}_k$ that repackages the contribution of a conformal family,
\begin{equation}
\mathcal{O}_i(x_i)\mathcal{O}_j(x_j)=x^{-2\Delta}_{ij}\sum\limits_kC_{ijk}{\mathcal{B}}^{ij}_k(x_i,x_j),\qquad x_{ij}=|x_i-x_j|\, .
\end{equation}

Kinematic space has been defined as the space of pairs of CFT points, or equivalently as the space of boundary anchored geodesics in pure AdS \cite{Czech2016,Boer2016}. Since OPE blocks are functions of two boundary points they are fields on kinematic space, and this suggests that they are related to the geodesics of the bulk dual. Indeed, it was shown that for pure AdS the dual of a scalar OPE block is a bulk field integrated over a boundary anchored geodesic,
\begin{equation}\label{Duality}
{\mathcal{B}}^{ij}_k(x_i,x_j)\sim\int_{\gamma_{ij}}ds\ \phi_k(x)\,,
\end{equation}
where $\gamma_{ij}$ is the geodesic with endpoints $(x_i,x_j)$ and $\phi_k$ is the scalar field dual to  $\mathcal{O}_k$. 

The duality between the OPE blocks and geodesic integrated fields was established by showing that both objects behave as fields on kinematic space with the same equation of motion, and the same boundary conditions. Each OPE block built from a scalar quasiprimary $\mathcal{O}_k$ is in an irreducible representation of the conformal group and satisfies an eigenvalue equation under the action of a quadratic conformal Casimir $L^2$, with the eigenvalue induced from $\mathcal{O}_k$,
\begin{equation}
[ L^2,{\mathcal{B}}^{ij}_k(x_i,x_j)]=-\Delta_k(\Delta_k-2){\mathcal{B}}^{ij}_k(x_i,x_j)\,.
\end{equation}
 By expressing the Casimir operator in the differential representation appropriate for ${\mathcal{B}}^{ij}_k$, this becomes a Laplacian on the dS$_2\ \times$ dS$_2$ kinematic space, 
 \begin{equation}\label{OPEblockEoM}
2[\Box_{\text{dS}_2}+\bar{\Box}_{\text{dS}_2}]{\mathcal{B}}^{ij}_k(x_i,x_j)=-\Delta_k(\Delta_k-2){\mathcal{B}}^{ij}_k(x_i,x_j)\,.
\end{equation}

On the other hand, the bulk scalar field $\phi_k(x)$ dual to $\mathcal{O}_k$ satisfies a wave equation on AdS$_3$, with its mass related to $\Delta_k$ by the holographic dictionary,
\begin{equation}
\Box_{\text{AdS}_3} \phi_k(x)=m^2\phi_k(x)=\Delta_k(\Delta_k-2)\phi_k(x)\,.
\end{equation}
Then, the remarkable intertwining property of isometry generators determines the equation of motion for the geodesic integrated field \cite{Czech2016}
\begin{equation}
\int_{\gamma_{ij}}ds\ \Box_{AdS_3} \phi_k(x)=-2[\Box_{\text{dS}_2}+\bar{\Box}_{\text{dS}_2}]\int_{\gamma_{ij}}ds \ \phi_k(x)\,.
\end{equation}
The conclusion is that the geodesic integrated field obeys the same kinematic space wave equation \eqref{OPEblockEoM} as the OPE block,
\begin{equation}
2[\Box_{\text{dS}_2}+\bar{\Box}_{\text{dS}_2}]\int_{\gamma_{ij}}ds\ \phi_k(x)=-\Delta_k(\Delta_k-2)\int_{\gamma_{ij}}ds\ \phi_k(x)\,.
\end{equation}
Rounding out the proof requires showing both quantities satisfy the same constraints and the same boundary conditions, which determine the relative normalization omitted in \eqref{Duality}.

For pure AdS there is a one-to-one correspondence between pairs of spacelike separated boundary points and geodesics in the bulk. This makes it simple to identify both the space of pairs of boundary points, and the space of bulk geodesics as the same kinematic space. But for spacetimes that are locally AdS${}_3$, the existence of non-minimal geodesics in the bulk obfuscates this prescription. In such cases it is not \emph{a priori} clear in what sense the duality \eqref{Duality} holds.

This question was addressed for the case of conical defect spacetimes in \cite{Cresswell2017}. Static conical defects are locally AdS$_3$ geometries obtained from AdS$_3$ by a $\mathbb{Z}_{N}$ quotient in the angular direction, leaving a $2\pi/N$ periodic $\tilde \phi$ coordinate. This coordinate parametrizes the one dimensional boundary of a timeslice on which the OPE can be studied. The exact CFT states dual to the conical defect geometries will depend on the system under scrutiny, but in general they can be viewed as the CFT vacuum excited by a heavy operator that sources the defect in the bulk \cite{Balasubramanian1999,Balasubramanian2001,Lunin2002}. In the presence of other operators the OPE does not have an infinite radius of convergence, and it becomes more difficult to study the properties of the OPE blocks directly. Instead, in \cite{Cresswell2017} the excited CFT states were lifted to vacuum states of a covering space CFT on an $N$-times longer circle parametrized by $\phi$ \cite{Balasubramanian2015}. This process can be seen as removing the discrete $\mathbb{Z}_N$ symmetry of the base CFT states; only appropriately symmetrized quantities on the cover descend to observables on the base \cite{Balasubramanian2016}. 

With this construction, the OPE blocks in the base and cover CFTs can be related. Individual OPE blocks on the cover ${\mathcal{B}}_k(\phi_1, \phi_2)$ are not $\mathbb{Z}_N$ symmetric, but can be combined into gauge invariant observables dubbed partial OPE blocks,
\begin{equation}
{\mathcal{B}}_{k,m}(\alpha_m,\theta)=\frac{1}{N}|2-2\cos(2\alpha_m)|^{-\Delta_k}\sum\limits^{N-1}_{b=0}\exp\left(i\frac{2\pi b}{N}\frac{\partial}{\partial\theta}\right){\mathcal{B}}_k(\alpha_m,\theta)\,.
\end{equation}
 Here, the cover OPE blocks are written in terms of the half opening angle $\alpha=({\phi}_1-{\phi}_2)/2$ and centre angle $\theta=({\phi}_1+{\phi}_2)/2$. The angular distance $\alpha$ between operators is taken to be fixed at $\alpha_m$ while the rotations generated by $\partial/\partial \theta$ implement the symmetrization. The full OPE blocks in the base theory ${\mathcal{B}}'_k$ receive contributions from partial OPE blocks at all allowed angular separations $\alpha_m$ on the cover
\begin{equation}\label{OPEblocksCD-JAW}
{\mathcal{B}}'_k(\alpha,\theta)=\frac{1}{N}\sum\limits_{m=0}^{N-1}\exp\left(i\frac{2\pi m}{N}\frac{\partial}{\partial\phi_1}\right){\mathcal{B}}_{k,m}(\alpha_m,\theta)\,,
\end{equation}
where $\partial/\partial\phi_1$ generates changes in separation. 

Finally, it was shown that the partial OPE blocks individually satisfy duality relations like \eqref{Duality} as fields integrated over minimal or non-minimal geodesics in the conical defect spacetime. The angular separation $\alpha_m$ of the block ${\mathcal{B}}_{k,m}$ is related to the winding number of the geodesic in $\int_{\gamma_m} ds\ \phi_k$. Hence, the new observables ${\mathcal{B}}_{k,m}$ allow us to obtain more fine-grained information about the bulk spacetime that reaches beyond the entanglement shadow limiting minimal geodesics and Ryu-Takayanagi entanglement entropy.

Our approach in this paper will be similar, but can more readily be applied to the broad class of AdS$_3$ quotient geometries. We will argue that the base OPE blocks for states dual to these geometries can be obtained through the coordinate maps we develop as a sum over partial OPE blocks. The partial blocks are constructed to be invariant under the quotient action. We propose that a partial block is dual to a bulk field integrated over an individual geodesic, which can be minimal or not, as specified by the monodromy under the map. To avoid branch cuts in the full OPE blocks, we identify them as a sum over partial OPE blocks. 

While the bulk interpretation of the partial blocks is clear, they give the contribution to the OPE from individual geodesics or saddlepoints of the path length action \cite{Balasubramanian1999}, our new method also affords a better understanding of the CFT interpretation. Each partial block gives a contribution to the OPE as distinguished by the monodromy around the excited state's heavy operator insertion. To reach these results, we must first develop exact mappings between AdS$_3$ and the quotient geometries that can be used to transform the OPE blocks. We proceed with the Euclidean and Lorentzian cases in turn.

\subsection{AdS${}_3$ quotients}\label{orbifoldsec}
\subsubsection{Euclidean AdS}

One construction of AdS$_3$ is through the $\mathbb{R}^{3,1}$ embedding space. We start with the metric $ds^2=dX_0^2+dX_1^2+dX_2^2-dX_3^2$, with AdS$_3$ defined as the surface $X^2=X_0^2+X^2_1+X_2^2-X_3^2=-\ell^2$. There are a number of different parametrizations of this hyperboloid which give different patches of AdS. We focus on the Poincar\'e patch, which only covers part of the hyperboloid. To get the Poincar\'e metric, we implement the coordinates
\begin{gather}\label{EuclideanPoincareEmbeddingCoords}
      \begin{aligned}
         X_0&=\frac{1}{2u}\big(u^2-\ell^2+x^2+t^2\big)\\
	X_1&=\ell \,\frac{x}{u}\\
	X_2&=\ell \,\frac{t}{u}\\
	X_3&=\frac{1}{2u}\big(u^2+\ell^2+x^2+t^2\big) \,,
      \end{aligned}
\end{gather}
which leads to 
\begin{equation}
ds^2=\frac{\ell^2}{u^2}\big(dt^2+dx^2+du^2\big) \,.
\end{equation}
We can do a further coordinate transformation by setting $w=x+it$, $\bar w=x-it$, which gives us the metric
\begin{equation}\label{HartmanMetric}
ds^2=\frac{\ell^2}{u^2}\big(dw\,d\bar{w}+du^2\big)\,.
\end{equation}

Boundary anchored geodesics, and especially their lengths, will be very important for understanding the OPE block duality. In Poincar\'e coordinates, the geodesic distance $d$ along the embedding surface between two points $P_1$ and $P_2$ obeys
\begin{gather}
\begin{aligned}\label{generalgdlength}
\cosh \frac{d}{\ell}=-\frac{P_1\cdot P_2}{\ell^2}&=\frac{1}{2u_1 u_2}\left((t_1-t_2)^2+(x_1-x_2)^2+u_1^2+u_2^2\right)\\
&=\frac{1}{2u_1u_2}\left((w_1-w_2)(\bar{w}_1-\bar{w}_2)+u_1^2+u_2^2\right)\,.
\end{aligned}
\end{gather}
 In the limit where both points approach the boundary, such that $u_1,u_2\to0$ with their ratio held fixed, $u_1/u_2\rightarrow 1$, this becomes
\begin{align}
\cosh \frac{d}{\ell}&=1+\frac{1}{2u_1u_2}(w_1-w_2)(\bar w_1-\bar w_2) \,.
\end{align}
The length of a boundary anchored geodesic can then be approximated by
\begin{equation}\label{gdlength}
d\approx\ell\,\log\left(\frac{(w_1-w_2)(\bar w_1-\bar w_2)}{u_1u_2}\right)\,.
\end{equation}

We can also construct Poincar\'e AdS${}_3$ as a group manifold \cite{Banados1993}. This is done by considering each point $g$ in Euclidean AdS${}_3$ as an element of $\text{SL}(2,\mathbb{C})/\text{SU}(2)$ where, in the embedding coordinates,
\begin{equation}\label{EuclideanEmbeddingGroupElement}
g=\left(\begin{array}{cc}
X_3+X_0 & X_1+iX_2\\
X_1-iX_2 & X_3-X_0
\end{array}\right)\,.
\end{equation} 
For the Euclidean Poincar\'e embedding we have,
\begin{equation}
g=\left(\begin{array}{cc}
u+w\bar{w}/u & \ell w/u\\
\ell\bar{w}/u & \ell^2/u
\end{array}\right)\,.
\end{equation} 
The metric on AdS$_3$ (\ref{HartmanMetric}) is then given by the Cartan-Killing metric $ds^2=\frac{1}{2}\text{Tr}(g^{-1}dgg^{-1}dg)$ which has the correct isometry group for Poincar\'e AdS$_3$, SL$(2,\mathbb{C})/\mathbb{Z}_2$ \cite{Carlip1995}. Other locally AdS${}_3$ solutions are constructed as quotients by a subgroup of the isometry group. The subgroups we study in this paper are conjugacy classes generated by the elliptic, parabolic, and hyperbolic elements of the form
\begin{equation}\label{conjugacyclasses}
h_{\text{ell}}=\left(\begin{array}{cc}
e^{-i\pi\gamma} & 0\\
0 & e^{i\pi\gamma}
\end{array}\right),\qquad
h_{\text{para}}=\left(\begin{array}{cc}
1 & \alpha\\
0 & 1
\end{array}\right),\qquad
h_{\text{hyper}}=\left(\begin{array}{cc}
e^{\beta/2} & 0\\
0 & e^{-\beta/2}
\end{array}\right)\,,
\end{equation}
where $0<\gamma<1$, $\alpha\in\mathbb{C}$, and $\beta\in\mathbb{R}$. In each case, elements related by conjugation, $g\sim h g h^\dag$, are identified to obtain the quotient manifold. 

Each type of element produces a different locally AdS$_3$ solution. Identification using the elliptic element will give the conical defect, abbreviated `CD', with deficit angle $2\pi(1-\gamma)$. Accounting for the $\mathbb{Z}_2$ quotient of the isometry group, the subgroup generated by an elliptic element is the cyclic group $\mathbb{Z}_N$, where we take $N=1/\gamma\in \mathbb{N}$. The other two elements lead to infinite discrete groups. A quotient using the parabolic element with $\alpha=2\pi $ yields the massless BTZ black hole, which we abbreviate as `0M'. The hyperbolic element with $\beta=2\pi\sqrt{M}$ gives the static BTZ black hole with mass $M$, which we abbreviate as `BTZ'. In summary, the three types of quotient lead to identifications on the Poincar\'e patch as follows,
\begin{align}\label{CDident}
\text{CD:}\quad  & (w,u)\sim(e^{-2\pi i/N}w,u) \,,\\
\label{M=0ident}
\text{0M:}\quad & (w,u)\sim(w+2\pi \ell,u) \,,\\
\label{BTZident}
\text{BTZ:}\quad & (w,u)\sim(e^{2\pi\sqrt{M}}w,e^{2\pi\sqrt{M}}u) \,.
\end{align}
The $N \to\infty$ limit of the CD metric and the $M\to0$ limit of the BTZ metric both produce the 0M metric, but the respective conjugacy classes \eqref{conjugacyclasses} by which elements are identified are not related in this way. Some differences between these limits have been noted in \cite{Chen2018}. For these reasons we treat the 0M solution as a distinct case throughout.

Other locally AdS$_3$ solutions can be obtained using quotients by more complicated subgroups, such as a rotating BTZ black hole using a combination of elliptic and hyperbolic identifications, but we focus on the three archetypal examples above.

Finally, one may wonder if we can consider conical defects where $N>1$ but not an integer. Considering the rational case of $\gamma=m/n$, we find that the subgroup generated by this is $\mathbb{Z}_n$, which is not distinguishable from the integer case. For non-rational $\gamma$ things are worse, as the subgroup generated is no longer finite and the identification one gets is ambiguous. In addition, the validity of non-integer conical defects is suspect in string theory \cite{Lunin2002,Alday2006}, so we will not consider them further.

\subsubsection{Lorentzian AdS}

The Lorentzian case presents a challenge in our approach because the boundary cannot be described by a single complex coordinate. Still, one direct way of approaching Lorentzian AdS using our knowledge of the Euclidean case is to compare them on a timeslice. The $t=0$ slice in embedding coordinates is
\begin{gather}\label{LorentzianPoincareEmbeddingCoords}
      \begin{aligned}
         X_0&=\frac{1}{2u}\big(u^2-\ell^2+x^2\big)\\
	X_1&=\ell \,\frac{x}{u}\\
	X_2&=0\\
	X_3&=\frac{1}{2u}\big(u^2+\ell^2+x^2\big) \,.
      \end{aligned}
\end{gather}
This now satisfies the Lorentzian constraint equation $X^2=X_0^2+X^2_1-X_2^2-X_3^2=-\ell^2$ as well as the Euclidean one, allowing for direct comparison between signatures. On the timeslice the metric is
\begin{equation}
ds^2=\frac{\ell^2(dx^2+du^2)}{u^2}\,,
\end{equation}
which transforms to the upper half plane (UHP) using $s=x+iu$, $\bar{s}=x-iu$
\begin{equation}
ds^2=\frac{-4\ell^2dsd\bar{s}}{(s-\bar{s})^2}\,.
\end{equation}
The upper half plane inherits a $\text{PSL}(2,\mathbb{R})$ isometry group from the full $\text{SL}(2,\mathbb{R})\times \text{SL}(2,\mathbb{R})/\mathbb{Z}_2$ of Lorentzian AdS$_3$ when restricted to the timeslice. Again, we can describe a point $g$ in the timeslice using 
\begin{equation}\label{LorentzianEmbeddingGroupElement}
g=\left(\begin{array}{cc}
X_3+X_0 & X_1-X_2\\
X_1+X_2 & X_3-X_0
\end{array}\right)\,.
\end{equation} 
Then in the group manifold description a point on the UHP is
\begin{equation}
g=\frac{2 i}{s-\bar{s}}\left(\begin{array}{cc}
|s|^2 & \frac{s+\bar{s}}{2}\\
\frac{s+\bar{s}}{2} & 1
\end{array}\right)\,.
\end{equation}
The action of a $\text{PSL}(2,\mathbb{R})$ isometry group element
\begin{equation}
\left(\begin{array}{cc}
a & b\\
c & d
\end{array}\right)\,,\quad ad-bc=1\,,
\end{equation}
will transform the UHP coordinate as
\begin{equation}
s\rightarrow\frac{as+b}{cs+d}\,.
\end{equation}

The $\text{PSL}(2,\mathbb{R})$ isometry group also has the three different types of elements that define conjugacy classes. The elliptic, parabolic, and hyperbolic elements are now given by
\begin{equation}
h_{\text{ell}}=\left(\begin{array}{cc}
\cos\theta & -\sin\theta\\
\sin\theta & \cos\theta
\end{array}\right),\qquad
h_{\text{para}}=\left(\begin{array}{cc}
1 & \alpha\\
0 & 1
\end{array}\right),\qquad
h_{\text{hyper}}=\left(\begin{array}{cc}
e^{\beta/2} & 0\\
0 & e^{-\beta/2}
\end{array}\right)\,,
\end{equation}
where $0<\theta<2\pi$, $\alpha\in\mathbb{R}$, and $\beta\in\mathbb{R}$. Note that there are differences from the $\text{SL}(2,\mathbb{C})/\mathbb{Z}_2$ cases we had previously. In particular, the parabolic element involves a real value and the structure of the elliptic element is different. Once again, locally AdS$_3$ spacetimes are obtained as a quotient of $\text{PSL}(2,\mathbb{R})$ by subgroups. For the two BTZ cases, the identifications are exactly the same as before with $\alpha=2\pi$ and $\beta=2\pi \sqrt{M}$,
\begin{align}
&\text{0M:}\quad(x,u)\sim(x+2\pi \ell,u),\label{Lor0MIdentification}\\
&\text{BTZ:}\quad(x,u)\sim(e^{2\pi\sqrt{M}}x,e^{2\pi\sqrt{M}}u)\,.\label{LorBTZIdentification}
\end{align}
However, the identification in the elliptic case is significantly more complicated. We take $\theta=\pi/N$ with $N\in\mathbb{N}$ to reproduce the conical defect geometry, and find the identifications
 \begin{align}\label{LorCDIdentification}
\text{CD:}\quad&x\sim\frac{ \ell^2 x \cos(2\pi/N)+\tfrac{\ell}{2}(u^2+x^2-\ell^2)\sin(2\pi/N)}{\ell^2\cos^2(\pi/N)+\ell x\sin(2\pi/N)+(u^2+x^2)\sin^2(\pi/N)}\\
&u\sim\frac{\ell^2 u}{\ell^2\cos^2(\pi/N)+\ell x\sin(2\pi/N)+(u^2+x^2)\sin^2(\pi/N)}\,.
\end{align}
It is simpler in this case to use the complex $s$ coordinate, $s=x+iu$, which is identified as 
\begin{equation}\label{LorCDIdentificationS}
\text{CD:}\quad s\sim\frac{\ell \cos(\pi/N)s-\ell^2 \sin(\pi/N)}{\sin(\pi/N)s+\ell \cos(\pi/N)}\,.
\end{equation}

\subsection{AdS${}_3$ maps and metrics}\label{mapsec}
\subsubsection{Euclidean AdS}

We will be making use of powerful maps that relate pure AdS${}_3$ to other locally AdS${}_3$ geometries \cite{Banados1999a,Roberts2012}. We begin by considering a general AdS${}_3$ solution, written as
\begin{equation}\label{RobertsMetric}
ds^2=\ell^2 \left(-\frac{L}{2}dz^2-\frac{\bar{L}}{2} d\bar{z}^2+\left(\frac{1}{y^2}+\frac{y^2}{16}L\bar{L}\right) dz d\bar{z}+ \frac{dy^2}{y^2}\right)\,.
\end{equation}
We can see that for $L=\bar{L}=0$ this is the usual Poincar\'e metric of pure AdS${}_3$. More generally, we have the relationship 
\begin{equation}
T(z)=\frac{c}{12} L(z)\,,
\end{equation}
where $T(z)$ is the holomorphic stress tensor and $c=3\ell/2G$ is the usual central charge given by the Brown-Henneaux formula \cite{Brown1986}. The analogous relation holds for the anti-holomorphic stress tensor. For what follows, we will set $\ell=1$. 

The transformation of the stress tensor can be exploited to find maps between AdS$_3$ and the quotients. We consider starting with the usual Poincar\'e metric \eqref{HartmanMetric} and implementing the asymptotic relationship $w=f(z)$. The stress tensor transforms as
\begin{equation}
T(z)=\left(\frac{df}{dz}\right)^2T(w)+\frac{c}{12}\{f(z),z\}\,,
\end{equation}
where $\{f(z),z\}$ is the Schwarzian derivative. Since $T(w)=0$ for pure AdS, in the general spacetime \eqref{RobertsMetric} we have
\begin{equation}\label{L=Schw}
L(z)=\{f(z),z\} \,.
\end{equation}
From the CFT point of view, this allows us to get to any background we wish by identifying $f(z)$. Suppose we have the state $|\psi\rangle$ which is excited by an operator with weight $h_{\psi}$. Since
\begin{equation}
\langle\psi|T(z)|\psi\rangle = \frac{h_{\psi}}{z^2}\,,
\end{equation}
we can find the asymptotic map $f(z)$ relating this background to the flat background by solving the differential equation
\begin{equation}\label{hpsi}
\frac{h_{\psi}}{z^2}=\frac{c}{12}\{f(z),z\}\,.
\end{equation}

In turn, the asymptotic map $f(z)$ can be extended into the bulk using \cite{Roberts2012}
\begin{align}
\begin{aligned}\label{fullmaps}
w&=f(z)-\frac{2y^2f'(z)^2\bar{f}''(\bar{z})}{4f'(z)\bar{f}'(\bar{z})+y^2f''(z)\bar{f}''(\bar{z})}\,,\\
\bar{w}&=\bar{f}(\bar{z})-\frac{2y^2\bar{f}'(\bar{z})^2f''(z)}{4f'(z)\bar{f}'(\bar{z})+y^2f''(z)\bar{f}''(\bar{z})}\,,\\
u&=y\,\frac{4(f'(z)\bar{f}'(\bar{z}))^{3/2}}{4f'(z)\bar{f}'(\bar{z})+y^2f''(z)\bar{f}''(\bar{z})}\,,
\end{aligned}
\end{align}
which gives the full map between \eqref{HartmanMetric} and \eqref{RobertsMetric}. In addition, if there is a map $w=f(z)$ that asymptotically implements the transformation, then for any constants $a_1,a_2,a_3$, a more general solution to $\eqref{hpsi}$ is
\begin{equation}
\frac{a_1 f(z)}{1+a_2 f(z)}+a_3\,,
\end{equation}
which comes from $\text{SL}(2,\mathbb{C})$ invariance. These maps will give the same metric regardless of the $a_i$ parameters but the corresponding coordinate transformations will differ. For simplicity we take $a_1=1$, $a_2=a_3=0$. 

With this in place, we would like to work out the maps \eqref{fullmaps} for our AdS${}_3$ quotients. The three cases we study correspond in the CFT to states excited by operators with weights 
\begin{gather}
\begin{aligned}\label{3weights}
h_{\text{CD}}&=\frac{c}{24}\left(1-\frac{1}{N^2}\right)\,,\\
h_{\text{0M}}&=\frac{c}{24}\,,\\
h_{\text{BTZ}}&=\frac{c}{24}\big(1+M\big)\,.
\end{aligned}
\end{gather}
In the case of the conical defect, we can see the weight is that of the twist operator and these maps have been looked at before in other contexts \cite{Asplund2015,Anand2018}. The 0M case is the $N\rightarrow\infty$ or $M\rightarrow 0$ limit of the other two. Furthermore, these weights are all non-negative for $N\geq 1$ and $M\geq 0$, as they should be in a unitary CFT. 

These three cases lead to three differential equations \eqref{hpsi}. One can try to solve them using normal methods, or alternatively, one can surmise the form of $f(z)$ from invariance under the identifications \eqref{CDident},\ \eqref{M=0ident},\ \eqref{BTZident} found from the group manifold approach. These identifications suggest the asymptotic maps 
\begin{align}\label{CDasympMap}
f_{\text{CD}}(z) &= z^{-1/N}\,,\\\label{M=0asympMap}
f_{\text{0M}}(z)&=-i\log(z)\,,\\\label{BTZasympMap}
f_{\text{BTZ}}(z)&=\exp{\left(-i\sqrt{M}\log z\right)}\,,
\end{align}
which reproduce the expected weights. As can be seen from the form of the conjugacy classes \eqref{conjugacyclasses}, the $N\to\infty$ and $M\to 0$ limits produce the identity map, rather than the appropriate 0M map, further emphasizing its distinct character. 

Each asymptotic map can be extended into the bulk using \eqref{fullmaps}, which for the conical defect yields the full coordinate transformations
\begin{gather}
\begin{aligned}\label{CDFullTransformation}
w_{\text{CD}}&=\frac{z^{-1/N}((N^2-1)y^2+4N^2z\bar{z})}{((N+1)^2y^2+4N^2z\bar{z})}\,,\\
\bar{w}_{\text{CD}}&=\frac{\bar{z}^{-1/N}((N^2-1)y^2+4N^2z\bar{z})}{((N+1)^2y^2+4N^2z\bar{z})}\,,\\
u_{\text{CD}}&=\frac{4Ny(z\bar{z})^{(N-1)/2N}}{((N+1)^2y^2+4N^2z\bar{z})} \,.
\end{aligned}
\end{gather}
Similarly, for massless BTZ we have the full coordinate transformations
\begin{gather}
\begin{aligned}\label{M=0FullTransformation}
w_{\text{0M}}&=-i\frac{2y^2+(y^2+4z\bar{z})\log z}{y^2+4z\bar{z}}\,,\\
\bar{w}_{\text{0M}}&=i\frac{2y^2+(y^2+4z\bar{z})\log\bar{z}}{y^2+4z\bar{z}}\,,\\
u_{\text{0M}}&=\frac{4y\sqrt{z\bar{z}}}{y^2+4z\bar{z}} \,.
\end{aligned}
\end{gather}
Finally, for massive BTZ the full coordinate transformations are
\begin{gather}
\begin{aligned}\label{BTZFullTransformation}
w_{\text{BTZ}}&=\frac{\big((1-i\sqrt{M})^2y^2+4z\bar{z}\big) \exp{\big(-i\sqrt{M}\log z\big)}}{(1+M)y^2+4z\bar{z}}\,,\\
\bar{w}_{\text{BTZ}}&=\frac{\big((1+i\sqrt{M})^2y^2+4z\bar{z}\big) \exp{\big(i\sqrt{M}\log \bar{z}\big)}}{(1+M)y^2+4z\bar{z}}\,,\\
u_{\text{BTZ}}&=\frac{4y\sqrt{Mz\bar{z}}\,\exp\left(-\frac{i\sqrt{M}}{2}\log (\frac{z}{\bar{z}})\right) }{(1+M)y^2+4z\bar{z}}\,.
\end{aligned}
\end{gather}
Applying these transformations to pure AdS$_3$ yields metrics of the form \eqref{RobertsMetric}, with $L$ and $\bar L$ determined by \eqref{3weights} through \eqref{L=Schw} and \eqref{hpsi},
\begin{gather}
\begin{aligned}\label{3metrics}
ds^2_{\text{CD}}&=\frac{dzd\bar{z}+dy^2}{y^2}-\frac{1}{4}\big(1{-}\frac{1}{N^2}\big)\frac{dz^2}{z^2}-\frac{1}{4}\big(1{-}\frac{1}{N^2}\big)\frac{d\bar{z}^2}{\bar{z}^2}+\frac{1}{16}\big(1{-}\frac{1}{N^2}\big)^2\frac{y^2}{(z\bar{z})^2}dzd\bar{z}\,,\\
ds^2_{\text{0M}}&=\frac{dzd\bar{z}+dy^2}{y^2}-\frac{1}{4}\frac{dz^2}{z^2}-\frac{1}{4}\frac{d\bar{z}^2}{\bar{z}^2}+\frac{1}{16}\frac{y^2}{(z\bar{z})^2}dzd\bar{z}\,,\\
ds^2_{\text{BTZ}}&=\frac{dzd\bar{z}+dy^2}{y^2}-\frac{(1{+}M)}{4}\frac{dz^2}{z^2}-\frac{(1{+}M)}{4}\frac{d\bar{z}^2}{\bar{z}^2}+\frac{(1{+}M)^2}{16}\frac{y^2}{(z\bar{z})^2}dzd\bar{z}\,,
\end{aligned}
\end{gather}
which confirms that the asymptotic maps in \eqref{CDasympMap}-\eqref{BTZasympMap} produce the expected metrics when extended into the bulk. We finish by noting that although the massless BTZ metric can be obtained as a simple limit $N\rightarrow\infty$ or $M\rightarrow 0$ of the conical defect or BTZ metrics respectively, the coordinate transformations are not related in this way.

\subsubsection{Lorentzian AdS}\label{LorentzianMapSec}

The above maps do not generalize straightforwardly to the timeslice. However, we can again use the knowledge that the maps should respect the identifications \eqref{Lor0MIdentification}, \eqref{LorBTZIdentification}, and \eqref{LorCDIdentificationS} to determine 
\begin{align}
s_{\text{CD}}&= i\frac{1+z^{-1/N}}{1-z^{-1/N}}\,,\label{LorSTransformationCD}\\
s_{\text{0M}}&=-i\log(z)\,,\label{LorSTransformation0M}\\
s_{\text{BTZ}}&=\exp{\left(-i\sqrt{M}\log z\right)}\,\label{LorSTransformationBTZ}.
\end{align}
We note that these are full maps on the UHP, not asymptotic ones. The latter two are similar to the asymptotic maps we had before, as the identification on the timeslice is unaffected. The map for the conical defect has a similar piece, but needs to be changed to reflect the change in the elliptic element. In the following, it will be easiest to write the single complex coordinate $z$, which we will call the quotient coordinate for all three cases, as $z=re^{i\theta}$.

In the original $x,u$ coordinates, the map for the conical defect looks like
\begin{gather}\label{LorentzianCDtransform}
	\begin{aligned}
	x_{\text{CD}}&= \frac{2r^{-1/N}\sin(\theta/N)}{1+r^{-2/N}-2r^{-1/N}\cos(\theta/N)}\,,\\
	u_{\text{CD}}&=\frac{1-r^{-2/N}}{1+r^{-2/N}-2r^{-1/N}\cos(\theta/N)}\,.
	\end{aligned}
\end{gather}
For massless BTZ it takes the form
\begin{gather}\label{Lorentzian0Mtransform}
	\begin{aligned}
	x_{\text{0M}}&= \theta\,,\\
	u_{\text{0M}}&=-\log r\,.
	\end{aligned}
\end{gather}
Finally for massive BTZ it looks like
\begin{gather}\label{LorentzianBTZtransform}
	\begin{aligned}
	x_{\text{BTZ}}&=e^{\sqrt{M}\theta}\cos(\sqrt{M}\log r)\,,\\
	u_{\text{BTZ}}&=-e^{\sqrt{M}\theta}\sin(\sqrt{M}\log r)\,.
	\end{aligned}
\end{gather}
In the first two cases the boundary $u=0$ is when $r=1$ in the new coordinates, but for massive BTZ  we have two boundaries, $r=1$ and $r=\exp\left(-\frac{\pi}{\sqrt{M}}\right)$. The identification also produces a horizon at $x=0$ in the Poincar\'e coordinates which interpolates between the boundaries \cite{Fuente2014}. Furthermore, to have $u\geq0$, we need $r>1$ for CD, $r\leq 1$ for 0M, and $\exp\left(-\frac{\pi}{\sqrt{M}}\right)\leq r\leq 1$ for BTZ. Transforming the metric with these maps produces
\begin{align}
ds^2_{\text{CD}}&=\frac{4r^{2/N}}{N^2r^2(r^{2/N}-1)^2}(dr^2+r^2d\theta^2)\,,\\
ds^2_{\text{0M}}&=\frac{1}{r^2\log(r)^2}(dr^2+r^2d\theta^2)\,,\\
ds^2_{\text{BTZ}}&=\frac{M}{r^2\sin^2(\sqrt{M}\log r)}(dr^2+r^2d\theta^2)\,.
\end{align}
We see that the limits $N\to\infty$ and $M\to 0$ reproduce the 0M metric, while taking $N\to 1$ or inserting $M= -1$ gives back pure AdS$_3$. 

Finally, for the CFT analysis, we are interested in the asymptotic maps which are now easily obtained from the full ones
\begin{align}
x_{\text{CD}}&= \cot\left(\frac{\theta}{2N}\right)\,,\label{LorAsyMapCD}\\
x_{\text{0M}}&=\theta,\label{LorAsyMap0M}\\
x_{\text{BTZ}}&=\pm e^{\sqrt{M}\theta}\,.\label{LorAsyMapBTZ}
\end{align}
Note that the sign in the BTZ case will depend on which boundary one considers. We can interpolate between the two boundaries by analytic continuation, $\theta\rightarrow\theta+i\frac{\pi}{\sqrt{M}}$ \cite{Goto2017}. Further, if we interpret $\theta$ to be the complex angle of $z=re^{i\theta}$, the monodromy $z=ze^{2\pi i}$ will implement the identifications \eqref{LorCDIdentificationS}, \eqref{Lor0MIdentification}, and \eqref{LorBTZIdentification}, similarly to the Euclidean case.

\section{Bulk analysis of geodesic structure}\label{sec:3}
\subsection{Euclidean analysis}\label{EuclideanGeodesicSec}

In this section we use the maps between Poincar\'e AdS$_3$ and the quotient geometries to study the resulting structure of geodesics via the group manifold approach. The non-analyticities in the maps allow us to distinguish geodesics with different winding numbers.

Since the geometries \eqref{3metrics} are all locally AdS$_3$, the properties of their geodesics are closely related to those of pure AdS$_3$. More concretely, the lengths of quotient geodesics are given by lengths of AdS$_3$ geodesics whose endpoints are related by the quotient action. We calculate them using the method outlined in \cite{Maxfield2015}. We consider points $p,q$ in the group manifold of AdS$_3$ as in equation (\ref{EuclideanEmbeddingGroupElement}). The length of the geodesic between these points found in \eqref{generalgdlength} is then rewritten as
\begin{equation}
d(p,q)=\cosh^{-1}\left(\frac{\text{Tr}(p^{-1}q)}{2}\right)\,.
\end{equation}
The boundary is represented by singular matrices $p,q$, up to a divergent factor, and the geodesic distance between them diverges. We regulate by considering curves $p(\rho)$, $q(\rho)$ which approach the boundary as $\rho\to\infty$, and which have the property that $\lim_{\rho\to \infty}p(\rho)/\rho=p_{\partial}$, and similarly $q_\partial$, are finite and non-zero. Then in the boundary limit the geodesic length goes to
\begin{equation}\label{RegulatedGeoLength}
d(p_\partial,q_\partial)=\log\rho^2+\log(\text{Tr}(R_{\perp}p^T_{\partial}R_\perp^Tq_{\partial}))+O(1)\,,
\end{equation}
where $R_\perp=(\begin{smallmatrix}0&-1\\1&0\end{smallmatrix})$. The correction term indicates that any rescaling of $\rho$ can give a different finite contribution. In our quotient coordinates, we choose $\rho=1/\epsilon$ where the boundary is cut off at $y=\epsilon$. The radial coordinate is different for each of the different quotient geometries, so the different regulators are labelled.

This approach affords a very clear understanding of non-minimal geodesic lengths. We quotient the AdS$_3$ group manifold by the discrete group generated by one element from \eqref{conjugacyclasses}. The length of the geodesic connecting the boundary points $p_\partial$ and $h q_\partial h^\dag$ is still given by \eqref{RegulatedGeoLength}, 
\begin{equation}\label{NonMinGeoLength}
d(p_\partial,h q_\partial h^\dag)=\log\rho^2+\log(\text{Tr}(R_{\perp}p^T_{\partial}R_\perp^T h q_{\partial}h^\dag))+O(1)\,,
\end{equation}
but in the quotient spacetime $q_\partial$ and $ h q_\partial h^\dag$ are identified. Typically $d(p_\partial,q_\partial)\neq d(p_\partial,h q_\partial h^\dag)$. We now show that non-minimal geodesics can also be identified from monodromies in the asymptotic maps.

We now parametrize the points in the quotient manifold by mapping the embedding coordinates for Poincar\'e, equation (\ref{EuclideanPoincareEmbeddingCoords}), to our quotient coordinates $(z,\bar{z},y)$. Using (\ref{EuclideanEmbeddingGroupElement}) to find the group elements yields
\begin{align}
\text{CD: }&\frac{(z\bar{z})^{\frac{1-N}{2N}}}{4Ny}\left(\begin{array}{cc}
(z\bar{z})^{-\frac{1}{N}}((N-1)^2y^2+4N^2z\bar{z}) & z^{-\frac{1}{N}}((N^2-1)y^2+4N^2z\bar{z})\\
\bar{z}^{-\frac{1}{N}}((N^2-1)y^2+4N^2z\bar{z})& (N+1)^2y^2+4N^2z\bar{z}
\end{array}\right),\\
\text{0M: }&\frac{1}{4y\sqrt{z\bar{z}}}\left(\begin{array}{cc}
2y^2(2+\log(z\bar{z}))+(y^2+4z\bar{z})\log z\log\bar{z}& -i(2y^2+(y^2+4z\bar{z})\log z)\\
i(2y^2+(y^2+4z\bar{z})\log\bar{z})& y^2+4z\bar{z}
\end{array}\right),\\
\text{BTZ: }&\frac{z^{-(1-i\sqrt{M})/2}\bar{z}^{-(1+i\sqrt{M})/2}}{4\sqrt{M}y}\\\nonumber
&\times\left(\begin{array}{cc}
z^{-i\sqrt{M}}\bar{z}^{i\sqrt{M}}((M+1)y^2+4z\bar{z}) & z^{-i\sqrt{M}}((1-i\sqrt{M})^2y^2+4z\bar{z})\\
\bar{z}^{i\sqrt{M}}((1+i\sqrt{M})^2y^2+4z\bar{z})& (M+1)y^2+4z\bar{z}
\end{array}\right)\,.
\end{align}
One can check that conjugation by the elliptic, parabolic, or hyperbolic generators corresponds to taking $z\rightarrow ze^{2\pi i}$ for the respective points. To consider boundary points we take the limit described above resulting in
\begin{align}
\text{CD: }\qquad&\frac{N(z\bar{z})^{\frac{1+N}{2N}}}{\epsilon_{\text{CD}}}\left(\begin{array}{cc}
(z\bar{z})^{-\frac{1}{N}} & z^{-\frac{1}{N}}\\
\bar{z}^{-\frac{1}{N}}& 1
\end{array}\right),\\
\text{0M: }\qquad&\frac{\sqrt{z\bar{z}}}{\epsilon_{\text{0M}}}\left(\begin{array}{cc}
\log z\log\bar{z}& -i\log z\\
i\log\bar{z}& 1
\end{array}\right),\\
\text{BTZ: }\qquad&\frac{z^{(1+i\sqrt{M})/2}\bar{z}^{(1-i\sqrt{M})/2}}{\sqrt{M}\epsilon_{\text{BTZ}}}\left(\begin{array}{cc}
z^{-i\sqrt{M}}\bar{z}^{i\sqrt{M}} & z^{-i\sqrt{M}}\\
\bar{z}^{i\sqrt{M}}& 1
\end{array}\right)\,.
\end{align}
Now we can pick two points, say $z_1$ and $z_2$, and compute the geodesic length using equation (\ref{RegulatedGeoLength}),
\begin{align}\label{EucCDlength}
d_{\text{CD}}&=\log\left[N^2(z_1^{\frac{1}{N}}-z_2^{\frac{1}{N}})(\bar{z}_1^{\frac{1}{N}}-\bar{z}_2^{\frac{1}{N}})\right]+\frac{N-1}{2N}\log z_1\bar{z}_1z_2\bar{z}_2 -2\log \epsilon_{\text{CD}},\\\label{Euc0Mlength}
d_{\text{0M}}&=\log\left[(\log z_1-\log z_2)(\log\bar{z}_1-\log\bar{z}_2)\right]+\frac{1}{2}\log z_1\bar{z}_1z_2\bar{z}_2-2\log{\epsilon_{\text{0M}}},\\\label{EucBTZlength}
d_{\text{BTZ}}&=\log\left[M^{-1}(z_1^{i\sqrt{M}}-z_2^{i\sqrt{M}})(\bar{z}_2^{i\sqrt{M}}-\bar{z}_1^{i\sqrt{M}})\right]+\frac{1-i\sqrt{M}}{2}\log z_1\bar{z}_1z_2\bar{z}_2 -2\log \epsilon_{\text{BTZ}}\,.
\end{align}
Like for the metrics, but unlike for the transformations, the 0M geodesic distance is correctly obtained by taking either $N\to\infty$ or $M\to 0$. Since conjugation by a quotient generator takes $z\to z e^{2\pi i}$, and with reference to \eqref{NonMinGeoLength}, we also obtain winding geodesic lengths from these formulae. This demonstrates how non-analyticities in the asymptotic maps give rise to winding geodesics in the defect geometries.

\subsection{Lorentzian analysis}

We can proceed similarly using our maps (\ref{LorentzianCDtransform}), (\ref{Lorentzian0Mtransform}), and (\ref{LorentzianBTZtransform}) on the embedding coordinates (\ref{LorentzianPoincareEmbeddingCoords}) to find the matrix representations of points in the various defects as
\begin{align}
\text{CD: }\quad&\frac{1}{1-r^{-2/N}}\left(\begin{array}{cc}
1+r^{-2/N}+2r^{-1/N}\cos\left(\frac{\theta}{N}\right) & 2r^{-1/N}\sin\left(\frac{\theta}{N}\right)\\
2r^{-1/N}\sin\left(\frac{\theta}{N}\right)& 1+r^{-2/N}-2r^{-1/N}\cos\left(\frac{\theta}{N}\right) 
\end{array}\right),\\
\text{0M: }\quad&-\frac{1}{\log r}\left(\begin{array}{cc}
\theta^2+\log r ^2 & \theta\\
\theta& 1
\end{array}\right),\\
\text{BTZ: }\quad&-\frac{1}{\sin(\sqrt{M}\log r)}\left(\begin{array}{cc}
e^{\sqrt{M}\theta} & \cos(\sqrt{M}\log r)\\
\cos(\sqrt{M}\log r)& e^{-\sqrt{M}\theta}
\end{array}\right).
\end{align}
Again, conjugation by the appropriate quotient generator takes $\theta\rightarrow \theta+ 2\pi$. For boundary points we take the limit $r-1=\epsilon\rightarrow0$ in the conical defect case, and $1-r=\epsilon\rightarrow0$ in the massless and massive BTZ cases. This is due to the difference in domains of $r$, as described in Sec. \ref{LorentzianMapSec}. Taking these limits gives the points
\begin{align}
\text{CD: }\qquad&\frac{2N\sin^2\left(\frac{\theta}{2N}\right) }{\epsilon_{\text{CD}}}\left(\begin{array}{cc}
\cot^2\left(\frac{\theta}{2 N}\right) & \cot\left(\frac{\theta}{2N}\right)\\
\cot\left(\frac{\theta}{2N}\right)& 1
\end{array}\right),\\
\text{0M: }\qquad&\frac{1}{\epsilon_{\text{0M}}}\left(\begin{array}{cc}
\theta^2& \theta\\
\theta& 1
\end{array}\right),\\
\text{BTZ: }\qquad&\frac{e^{-\sqrt{M}\theta}}{\sqrt{M}\epsilon_{\text{BTZ}}}\left(\begin{array}{cc}
e^{2\sqrt{M}\theta} & e^{\sqrt{M}\theta}\\
e^{\sqrt{M}\theta}& 1
\end{array}\right).
\end{align}
We can pick two points on the boundary circle, $\theta_1$ and $\theta_2$, to find the geodesic lengths from \eqref{RegulatedGeoLength},
\begin{align}
d_{\text{CD}}&=\log\left[4N^2\sin^2\left(\frac{\theta_1-\theta_2}{2N}\right)\right]-2\log\epsilon_{\text{CD}}\,,\label{LorCDlength}\\
d_{\text{0M}}&=\log[(\theta_1-\theta_2)^2]-2\log\epsilon_{\text{0M}}\,,\label{Lor0Mlength}\\
d_{\text{BTZ}}&=\log\left[\frac{4}{M}\sinh^2\left(\sqrt{M}\frac{\theta_1-\theta_2}{2}\right)\right]-2\log\epsilon_{\text{BTZ}}\,.\label{LorBTZlength}
\end{align}
Again, we see a nice smooth limit between the $N\rightarrow\infty$ and $M\rightarrow0$ limits for the massless BTZ geodesic lengths even though their maps and their embedding coordinates do not have a smooth limit.

In the BTZ expression above we took both points to be on the same boundary $r=1$. Points on the $r=\exp[-\frac{\pi}{\sqrt{M}}]$ boundary are parametrized as
\begin{equation}
\text{BTZ: }\qquad\frac{e^{-\sqrt{M}\theta}}{\sqrt{M}\tilde{\epsilon}_{\text{BTZ}}}\left(\begin{array}{cc}
e^{2\sqrt{M}\theta} & -e^{\sqrt{M}\theta}\\
-e^{\sqrt{M}\theta}& 1
\end{array}\right)\,,
\end{equation}
where we have a different regulator, $\exp[\frac{\pi}{\sqrt{M}}]r-1=\tilde{\epsilon}_{\text{BTZ}}\rightarrow0$. For two points on the $r=\exp[-\frac{\pi}{\sqrt{M}}]$ boundary the distance formula is unchanged, but for horizon crossing geodesics between the two boundaries the lengths are
\begin{equation}
d_{\text{BTZ, crossing}}=\log\left[\frac{4}{M}\cosh^2\left(\sqrt{M}\frac{\theta_1-\theta_2}{2}\right)\right]-\log\epsilon_{\text{BTZ}}\tilde{\epsilon}_{\text{BTZ}}\,.\label{gdlengthsBTZCross}
\end{equation}
Note that this is related to the single sided geodesic length with $\theta\rightarrow\theta+\frac{i\pi}{\sqrt{M}}$.

Once again, in view of \eqref{NonMinGeoLength} and the fact that quotient generators take $\theta\to \theta+2\pi$ we find that non-analyticities in the maps between pure AdS$_3$ and the quotient geometries distinguish boundary anchored geodesics of different windings.

\section{CFT analysis of OPE blocks}\label{sec:4}
\subsection{Euclidean analysis}\label{sec:4.1}

In this section we argue that the non-analyticities in the asymptotic maps between pure AdS$_3$ and the quotient geometries which distinguish winding geodesics also distinguish quotient invariant contributions to OPE blocks. The terms in the OPE block decomposition are in correspondence with the winding geodesics, which suggests a dual relationship. 

We start by mapping vacuum OPE blocks to a non-trivial background using the asymptotic maps from our bulk analysis. Consider a transformation $x\rightarrow x'$ where
\begin{equation}
\Omega(x')=\det\left(\frac{\partial x'^{\mu}}{\partial x^{\nu}}\right)\,.
\end{equation}
An OPE block $B$ of scalar operators will in general transform as \cite{Czech2016}
\begin{equation}\label{OPEBlocktransform}
{{B}}^{ij}_k(x_i,x_j)=\left(\frac{\Omega(x'_i) }{\Omega(x'_j)}\right)^{\Delta_{ij}/2}{{B}}^{ij}_k(x'_i,x'_j)\,,
\end{equation}
where $\Delta_{ij}\equiv \Delta_i -\Delta_j$\,. For simplicity, we will set $\Delta_{ij}=0$. Now we apply \cref{CDasympMap,M=0asympMap,BTZasympMap} for the CD, 0M, and BTZ cases respectively which naively gives the transformation
\begin{equation}
{{B}}^{ij}_k(z_i,\bar{z}_i,z_j,\bar{z}_j) ={{B}}^{ij}_k(w_i,\bar{w}_i,w_j,\bar{w}_j)\,.
\end{equation}
However, we immediately see a problem. All of these maps have a branch cut as we take $z\rightarrow ze^{2\pi i}$, whereas the OPE block should be a single-valued observable. If we wish to remove branch cuts from the OPE block, we should instead consider 
\begin{align}
{\text{CD:}}\quad {{\mathcal{B}}}^{ij}_k(z_i,\bar{z}_i,z_j,\bar{z}_j)
&=\sum_{p_i,p_j}{\mathcal{B}}^{ij}_k(w_ie^{-\frac{2\pi i p_i}{N}},\bar{w}_ie^{\frac{2\pi i p_i}{N}},w_je^{-\frac{2\pi i p_j}{N}},\bar{w}_je^{\frac{2\pi i p_j}{N}})\,,\\
{\text{0M:}}\quad {{\mathcal{B}}}^{ij}_k(z_i,\bar{z}_i,z_j,\bar{z}_j)
&=\sum_{p_i,p_j}{\mathcal{B}}^{ij}_k(w_i+2\pi p_i,\bar{w}_i+2\pi p_i,w_j+2\pi p_j,\bar{w}_j+2\pi p_j)\,,\\
{\text{BTZ:}}\quad {{\mathcal{B}}}^{ij}_k(z_i,\bar{z}_i,z_j,\bar{z}_j) &=\sum_{p_i,p_j} {\mathcal{B}}^{ij}_k(w_ie^{2\pi p_i\sqrt{M}},\bar{w}_ie^{2\pi p_i\sqrt{M}},w_je^{2\pi p_j\sqrt{M}},\bar{w}_je^{2\pi p_j\sqrt{M}}) \,.
\end{align}
These are sums over pre-images of points identified under the maps. Alternatively, these sums can be argued for from the quotient identifications on pure AdS$_3$ in equations (\ref{CDident}), (\ref{M=0ident}), and (\ref{BTZident}) respectively as they are invariant under the boundary action of the quotient. This method of images has been used frequently for describing quotient invariant observables \cite{Balasubramanian1999,Balasubramanian2003,Arefeva2016}.

We now relate these images to geodesics. Fixing one of the points in the vacuum OPE block and taking images of the other point defines a sequence of different geodesics in the pure AdS$_3$ bulk. Under the quotient these all map to geodesics with the same endpoints, but differing by their winding. For conical defects the paper \cite{Cresswell2017} found that fields integrated on each of these winding geodesics have a dual description, the partial OPE block, summarized in equation (\ref{OPEblocksCD-JAW}). Similarly, we can reorganize the sums above, decomposing the full OPE blocks into distinct contributions labelled by $m$,
\begin{align}\label{OPEblocksCD}
{\text{CD: }} {\mathcal{B}}^{ij}_k(z_i,\bar{z}_i,z_j,\bar{z}_j)&=\sum_{m}{\mathcal{B}}^{ij}_{k,m}\left(w_ie^{-\frac{2\pi i m}{N}},\bar{w}_ie^{\frac{2\pi i m}{N}},w_je^{\frac{-2\pi i m}{N}},\bar{w}_je^{\frac{2\pi im}{N}}\right)\,, \\\label{OPEblocksM=0}
{\text{0M: }} {\mathcal{B}}^{ij}_k(z_i,\bar{z}_i,z_j,\bar{z}_j)&=\sum_{m}{\mathcal{B}}^{ij}_{k,m}\left(w_i+2\pi m,\bar{w}_i+2\pi m,w_j+2\pi m,\bar{w}_j+2\pi m\right)\,, \\\label{OPEblocksBTZ}
{\text{BTZ: }} {\mathcal{B}}^{ij}_k(z_i,\bar{z}_i,z_j,\bar{z}_j)&=\sum_{m}{\mathcal{B}}^{ij}_{k,m}\left(w_ie^{2\pi m\sqrt{M}},\bar{w}_ie^{2\pi m\sqrt{M}},w_je^{2\pi m\sqrt{M}},\bar{w}_je^{2\pi m\sqrt{M}}\right) \,,
\end{align}
where
\begin{align}\label{PartialOPEblocksCD}
{\text{CD:  }} {\mathcal{B}}^{ij}_{k,m}(w_i,\bar{w}_i,w_j,\bar{w}_j)&=\sum_{b}{\mathcal{B}}^{ij}_{k}\left(w_ie^{\frac{2\pi i (m-b)}{N}},\bar{w}_ie^{\frac{-2\pi i (m-b)}{N}},w_je^{-\frac{2\pi i b}{N}},\bar{w}_je^{\frac{2\pi ib}{N}}\right)\,, \\\label{PartialOPEblocksM=0}
{\text{0M:  }}{\mathcal{B}}^{ij}_{k,m}(w_i,\bar{w}_i,w_j,\bar{w}_j)&=\sum_{b}{\mathcal{B}}^{ij}_{k}\left(w_i{+}2\pi(b{-}m),\bar{w}_i{+}2\pi(b{-}m),w_j{+}2\pi b,\bar{w}_j{+}2\pi b\right)\,, \\
{\text{BTZ:  }}{\mathcal{B}}^{ij}_{k,m}(w_i,\bar{w}_i,w_j,\bar{w}_j)&=\sum_{b}{\mathcal{B}}^{ij}_{k}\left(w_ie^{2\pi \sqrt{M}(b-m)},\bar{w}_ie^{2\pi \sqrt{M}(b-m)},w_je^{2\pi \sqrt{M}b},\bar{w}_je^{2\pi \sqrt{M}b}\right)\label{PartialOPEblocksBTZ} \,.
\end{align}
Each of the new quantities ${\mathcal{B}}^{ij}_{k,m}$ is invariant under the appropriate quotient action on both coordinates $z_{i,j}$ sending $z\to z e^{2\pi i}$, meaning they are valid observables in the  quotient coordinates. This has been expressed before in terms of invariance under the CFT's discrete gauge symmetry that is induced by the quotient \cite{Balasubramanian2016, Cresswell2017}. 

Our suggestion is that each partial OPE block ${\mathcal{B}}^{ij}_{k,m}(w_i,\bar{w}_i,w_j,\bar{w}_j)$ is dual to the bulk field integrated over a geodesic with winding related to the label $m$. By construction, each partial OPE block depends on pairs of boundary points at a fixed separation determined by $m$. This can be seen from the geodesic distance formulae, \cref{EucCDlength,Euc0Mlength,EucBTZlength}, by acting with the quotient generator $b$ times on point $z_1$, and $b+m$ times on point $z_2$, as dictated by \cref{PartialOPEblocksCD,PartialOPEblocksM=0,PartialOPEblocksBTZ} and \cref{OPEblocksCD,OPEblocksM=0,OPEblocksBTZ}:
\begin{align}
\label{CDgdlength}
d_{\text{CD}}(m,b)=&\log\left[N^2(z_1^{\frac{1}{N}}-z_2^{\frac{1}{N}}e^{2\pi m i/N})(\bar{z}_1^{\frac{1}{N}}-\bar{z}_2^{\frac{1}{N}}e^{-2\pi m i/N})\right]\\
\nonumber&+\frac{N-1}{2N}\log z_1\bar{z}_1z_2\bar{z}_2 -2\log \epsilon_{\text{CD}}
,\\
\label{0Mgdlength}
d_{\text{0M}}(m,b)=&\log\left[(\log z_1-\log z_2-2\pi m i)(\log\bar{z}_1-\log\bar{z}_2+2 \pi m i)\right]\\
\nonumber&+\frac{1}{2}\log z_1\bar{z}_1z_2\bar{z}_2-2\log{\epsilon_{\text{0M}}},\\
\label{BTZgdlength}
d_{\text{BTZ}}(m,b)=&\log\left[M^{-1}(z_1^{i\sqrt{M}}-z_2^{i\sqrt{M}}e^{2\pi \sqrt{M}m})(\bar{z}_2^{i\sqrt{M}}-\bar{z}_1^{i\sqrt{M}}e^{-2\pi \sqrt{M}m})\right]\\
\nonumber&+\frac{1-i\sqrt{M}}{2}\log z_1\bar{z}_1z_2\bar{z}_2 -2\log \epsilon_{\text{BTZ}}.
\end{align}
In each case we find that all dependence on the $b$-sum index drops out. This means that each vacuum OPE block entering ${\mathcal{B}}^{ij}_{k,m}(w_i,\bar{w}_i,w_j,\bar{w}_j)$ defines an AdS$_3$ geodesic, all of which have the same length and become identified under the quotient. Hence, each ${\mathcal{B}}^{ij}_{k,m}(w_i,\bar{w}_i,w_j,\bar{w}_j)$ picks out a unique geodesic in the dual quotient geometry, with winding specified by $m$. Blocks with different $m$ are related by repeated action of the quotient generator on only one of the boundary points in the same way that geodesics with different windings are related, as seen in \eqref{NonMinGeoLength} and the results of Section \ref{sec:3}.

The new quantities ${\mathcal{B}}^{ij}_{k,m}$ are each defined as a sum over vacuum OPE blocks which are known to be convergent inside correlation functions \cite{Pappadopulo2012,Rychkov2016}, but any required normalization has been neglected above. For the conical defect \eqref{PartialOPEblocksCD}, the sum is finite and can be normalized as
\begin{equation}\label{normalizedCD}
{\text{CD: }}\ {\mathcal{B}}^{ij}_{k,m}(w_i,\bar{w}_i,w_j,\bar{w}_j)=\frac{1}{N}\sum_{b=0}^{N-1}{\mathcal{B}}^{ij}_{k}\left(w_ie^{\frac{2\pi i (m-b)}{N}},\bar{w}_ie^{\frac{-2\pi i (m-b)}{N}},w_je^{-\frac{2\pi i b}{N}},\bar{w}_je^{\frac{2\pi ib}{N}}\right).
\end{equation}
 The $b$-sum ensures that ${\mathcal{B}}^{ij}_{k,m}$ is quotient invariant, but does not alter the overall contribution to the OPE. This follows since the $N$ terms in the sum each give equivalent contributions due to conformal symmetry, or from bulk considerations due to the equality of geodesic distances discussed in the previous paragraph.

 For the massless and massive BTZ cases, the $b$-sums are infinite making the normalization appear ambiguous and bringing the convergence of the sum into question. However, we know that the OPE itself is convergent in CFTs, and our ${\mathcal{B}}^{ij}_{k,m}$ represents only a partial contribution to the full OPE. Again, although an infinite number of images are included to ensure invariance under the quotient, each image represents an equivalent contribution by symmetry. We can normalize the operators using a formal limit
\begin{align}
{\text{0M: }}& \ {\mathcal{B}}^{ij}_{k,m}=\lim_{N\to\infty} \frac{1}{2N{+}1}\sum_{b=-N}^{N}{\mathcal{B}}^{ij}_{k}\left(w_i{+}2\pi(b{-}m),\bar{w}_i{+}2\pi(b{-}m),w_j{+}2\pi b,\bar{w}_j{+}2\pi b\right),\label{normalized0M} \\
{\text{BTZ: }}& \ {\mathcal{B}}^{ij}_{k,m}=\lim_{N\to\infty} \frac{1}{2N{+}1}\sum_{b=-N}^{N}{\mathcal{B}}^{ij}_{k}\left(w_ie^{2\pi \sqrt{M}(b-m)},\bar{w}_ie^{2\pi \sqrt{M}(b-m)},w_je^{2\pi \sqrt{M}b},\bar{w}_je^{2\pi \sqrt{M}b}\right).\label{normalizedBTZ}
\end{align}

In contrast, the full OPE blocks in \cref{OPEblocksCD,OPEblocksM=0,OPEblocksBTZ} are not sums over equivalent contributions. By convention we can arrange for the $m=0$ block to correspond to the minimal operator separation, and hence the minimal bulk geodesic. All other $m\neq0$ blocks are subleading since they represent operators at greater separation in the vacuum where there are no complications from the presence of other operators. The fall off with distance can be seen explicitly in the smeared representation for vacuum OPE blocks \cite{Czech2016}. The conical defect sum is finite and can be normalized as in \eqref{normalizedCD}, whereas for the BTZ cases, we see from \eqref{0Mgdlength} and \eqref{BTZgdlength} that the operators become infinitely separated for large $|m|$, and their contribution becomes negligible. This is the mechanism by which similar applications of the method of images for conical defects and BTZ spacetimes produce finite correlators from infinite sums \cite{Keski1999,Balasubramanian1999,Balasubramanian2013}.


\subsection{Lorentzian analysis}

The Lorentzian case is slightly different because the boundary is not parametrized by a complex coordinate. Still, we can rely on invariance under the quotient action to guide us. OPE blocks in the quotient coordinate $\theta$ transform to vacuum OPE blocks using eq. \eqref{OPEBlocktransform} with the asymptotic maps \eqref{LorAsyMapCD}-\eqref{LorAsyMapBTZ}. For simplicity, we will specialize to $\Delta_i=\Delta_j$. Once again, these maps are not invariant under $\theta\to \theta+2\pi$ meaning there is an ambiguity in the transformation of the naive defect OPE blocks. To define single-valued OPE blocks we sum over images, ensuring consistency with the $u\to0$ boundary limits of \eqref{Lor0MIdentification}-\eqref{LorCDIdentification}.  We then have the following transformations for OPE blocks
\begin{align}
{\text{CD:}}\quad {{\mathcal{B}}}^{ij}_k(\theta_i,\theta_j)
&=\sum_{p_i,p_j}{\mathcal{B}}^{ij}_k\left(\frac{\cos(p_i\pi/N)x_i{-}\sin(p_i\pi/N)}{\sin(p_i\pi/N)x_i{+}\cos(p_i\pi/N)},\frac{\cos(p_j\pi/N)x_j{-}\sin(p_j\pi/N)}{\sin(p_j\pi/N)x_j{+}\cos(p_j\pi/N)}\right)\,,\\
{\text{0M:}}\quad {{\mathcal{B}}}^{ij}_k(\theta_i,\theta_j)
&=\sum_{p_i,p_j}{\mathcal{B}}^{ij}_k(x_i+2\pi p_i,x_j+2\pi p_j)\,,\\
{\text{BTZ:}}\quad {{\mathcal{B}}}^{ij}_k(\theta_i,\theta_j)&=\sum_{p_i,p_j} {\mathcal{B}}^{ij}_k\left(x_ie^{2\pi p_i\sqrt{M}},x_je^{2\pi p_j\sqrt{M}}\right) \,.
\end{align}
For the BTZ case we have written the single sided OPE block above. The OPE block relating operators on different boundaries is related by the analytic continuation of one of the $\theta$ coordinates,
\begin{align}
{\text{BTZ, crossing:}}\quad {{\mathcal{B}}}^{ij}_k(\theta_i+i\pi/\sqrt{M},\theta_j)&=\sum_{p_i,p_j} {\mathcal{B}}^{ij}_k\left(-x_ie^{2\pi p_i\sqrt{M}},x_je^{2\pi p_j\sqrt{M}}\right) \,.
\end{align}
This matches nicely with the analytic continuation found both in the coordinate transformations \eqref{LorAsyMapBTZ} and in the geodesic lengths \eqref{gdlengthsBTZCross}. 

As before we can reorganize the sums, writing them as a decomposition into quotient invariant partial OPE blocks 
\begin{align}
{\text{CD:}}&\quad {{\mathcal{B}}}^{ij}_k(\theta_i,\theta_j)
=\sum_{m}{\mathcal{B}}^{ij}_{k,m}\left(\frac{\cos(m\pi/N)x_i{-}\sin(m\pi/N)}{\sin(m\pi/N)x_i{+}\cos(m\pi/N)},\frac{\cos(m\pi/N)x_j{-}\sin(m\pi/N)}{\sin(m\pi/N)x_j{+}\cos(m\pi/N)}\right)\,,\\
{\text{0M:}}&\quad {{\mathcal{B}}}^{ij}_k(\theta_i,\theta_j)
=\sum_{m}{\mathcal{B}}^{ij}_{k,m}(x_i+2\pi m,x_j+2\pi m)\,,\\
{\text{BTZ:}}&\quad {{\mathcal{B}}}^{ij}_k(\theta_i,\theta_j)=\sum_{m} {\mathcal{B}}^{ij}_{k,m}\left(x_ie^{2\pi m\sqrt{M}},x_je^{2\pi m\sqrt{M}}\right) \,,\\
{\text{BTZ,}}&\text{ crossing:}\quad {{\mathcal{B}}}^{ij}_k\left(\theta_i+i\pi/\sqrt{M},\theta_j\right)=\sum_{m} {\mathcal{B}}^{ij}_{k,m}\left(-x_ie^{2\pi m\sqrt{M}},x_je^{2\pi m\sqrt{M}}\right) \,.
\end{align}
where 
\begin{align}
{\text{CD:}}&\quad {\mathcal{B}}^{ij}_{k,m}(x_i,x_j)\\\nonumber
&\quad=\sum_{b}{\mathcal{B}}^{ij}_{k}\left(\frac{\cos((b{-}m)\pi/N)x_i{-}\sin((b{-}m)\pi/N)}{\sin((b{-}m)\pi/N)x_i{+}\cos((b{-}m)\pi/N)},\frac{\cos(b\pi/N)x_j{-}\sin(b\pi/N)}{\sin(b\pi/N)x_j{+}\cos(b\pi/N)}\right)\,,\\
{\text{0M:}}&\quad {\mathcal{B}}^{ij}_{k,m}(x_i,x_j)
=\sum_{b}{\mathcal{B}}^{ij}_{k}(x_i+2\pi (b-m),x_j+2\pi b)\,,\\
{\text{BTZ:}}&\quad {\mathcal{B}}^{ij}_{k,m}(x_i,x_j)
=\sum_{b} {\mathcal{B}}^{ij}_{k}\left(x_ie^{2\pi (b-m)\sqrt{M}},x_je^{2\pi b\sqrt{M}}\right) \,.
\end{align}
For the BTZ partial OPE blocks, the above equation encompasses both signs of the $x$ coordinates allowed in  \eqref{LorAsyMapBTZ}. 

The partial OPE blocks $ {\mathcal{B}}^{ij}_{k,m}(x_i,x_j)$ give the contribution to the full OPE block from image operators at a fixed separation in $x$, indicated by the label $m$. Each vacuum OPE block included in the sum gives an identical contribution, as is apparent by the conformal symmetry of the vacuum state, but the sum is necessary for manifest invariance under the quotient. This can be compared with the geodesic distance formulae, \cref{LorCDlength,Lor0Mlength,LorBTZlength} and \eqref{gdlengthsBTZCross}. Acting with the quotient generator $b$ times on point $\theta_1$, and $b+m$ times on point $\theta_2$ gives

 \begin{align}
\label{LorCDlengthbm}
d_{\text{CD}}(b,m)&=\log\left[4N^2\sin^2\left(\frac{\theta_1-\theta_2-2\pi m}{2N}\right)\right]-2\log\epsilon_{\text{CD}}
,\\
\label{Lor0Mlengthbm}
d_{\text{0M}}(b,m)&=\log\left[(\theta_1-\theta_2+2\pi m)^2\right]-2\log \epsilon_{\text{0M}},\\
\label{LorBTZlengthbm}
d_{\text{BTZ}}(b,m)&=\log\left[\frac{4}{M}\sinh^2\left(\sqrt{M}\frac{\theta_1-\theta_2-2\pi m}{2}\right)\right]-2\log\epsilon_{\text{BTZ}},\\
\label{LorBTZcrossinglengthbm}
d_{\text{BTZ, crossing}}(b,m)&=\log\left[\frac{4}{M}\cosh^2\left(\sqrt{M}\frac{\theta_1-\theta_2-2\pi m}{2}\right)\right]-\log\epsilon_{\text{BTZ}}\tilde\epsilon_{\text{BTZ}}.
\end{align}
In every case the dependence on $b$ drops out, showing a precise matching between the behaviour of geodesics and the structure of ${\mathcal{B}}^{ij}_{k,m}(x_i,x_j)$. Since each term gives an equivalent contribution, the partial OPE blocks can be normalized in the same way as described in Section \ref{sec:4.1}.

Each ${\mathcal{B}}^{ij}_{k,m}(x_i,x_j)$ block is invariant when the quotient acts on both $x_{i,j}$, while blocks with different $m$ are related by repeated action on only one of $x_{i,j}$. Winding or crossing geodesics of different lengths are related by the repeated quotient action on one endpoint, and each is invariant under the action on both endpoints. Hence, we also interpret the $ {\mathcal{B}}^{ij}_{k,m}(x_i,x_j)$ as giving the contribution to the full OPE block from the dual bulk field integrated over a single geodesic, which may be minimal, winding, or horizon crossing as appropriate.

\section{Discussion}\label{sec:5}

In this paper we have explored generalizing the holographic duality between OPE blocks and geodesic integrated fields to non-trivial locally AdS${}_3$ spacetimes, both in the Euclidean case and for the Poincar\'e disk of Lorentzian AdS$_3$. Such spacetimes can be described as quotients of AdS$_3$ by discrete subgroups of the isometry group. We found that the transformations between AdS$_3$ and its quotients involve non-analyticities which lead to branch cuts in OPE blocks for the dual excited CFT states. We proposed that the branch cuts should be removed by summing over image points of the quotient action, while also noting a natural decomposition of the OPE blocks into quotient invariant contributions. These contributions, partial OPE blocks, are observables in and of themselves, carrying more fine-grained information than the full OPE block. We explained how this decomposition arises from the coordinate transformations, and offered a dual interpretation of the partial OPE blocks as bulk fields integrated over individual winding or crossing geodesics.

On the bulk side we presented coordinate transformations between pure AdS$_3$ and the conical defect, the massless BTZ black hole, and massive BTZ geometries. These maps incorporate the corresponding quotient identifications, which are expressed as a monodromy of the complex coordinate describing the defect spacetime. The identifications map sets of boundary anchored geodesics between distinct pairs of points in pure AdS$_3$ to geodesics with identical endpoints in the new spacetime, differentiated by their winding around the defect. We showed how the lengths of these geodesics transform emphasizing the relation to monodromy. 

In the CFT we showed that branch cuts appear in OPE blocks after the transformation from pure AdS$_3$ to the quotient spacetime. Removing these branch cuts by summing over images led to a new quotient invariant quantity, the partial OPE block. This process can also be seen as requiring the OPE blocks to be invariant under a discrete gauge symmetry induced by the quotient. The various partial OPE blocks are related by applying the quotient generator to one of the insertion points. The same action distinguishes geodesics with different winding. In view of the duality known for pure AdS$_3$, we conjecture that partial OPE blocks are dual to fields integrated over the individual geodesics in the bulk which can be minimal, non-minimal, or even horizon crossing. 

In the case of the conical defect, the discrete quotient group is finite and therefore isomorphic to $\mathbb{Z}_N$. However, for both BTZ cases, the group is infinite and the interpretation of how the orbifold CFT is properly defined is less clear. The idea of orbifolding by these infinite discrete groups is not new \cite{Horowitz1998}, but our interpretation of how these discrete gauge symmetries affect the OPE blocks and their dual is. We have not proven explicitly that the partial OPE blocks are dual to fields integrated over the minimal or non-minimal geodesics, as this would require a greater understanding of the intertwining relation for the Radon transform in non-pure AdS$_3$ \cite{Czech2016}. 

Differences arise between the Euclidean and Lorenztian descriptions for the obvious reason: the monodromy of the $z$ coordinate only exists if $z$ is complex. In Euclidean signature the boundary is naturally complex and the monodromy affecting OPE blocks is easily understood. In Lorentzian signature we restricted our considerations to the upper half plane description of the Poincar\'e disk to accord with this. In the full Lorentzian case, it is difficult to see how we could reduce the action of the quotient into the monodromy of a complex coordinate as it is unclear what the correct combination of coordinates would be. In addition, for Lorentzian AdS$_3$ there are no geodesics between timelike separated boundary points, whereas OPE blocks for timelike separated insertions remain well-defined. It would be interesting to understand the duality in these cases, and also to find maps analogous to those displayed here for coordinate systems other than Poincar\'e, in both the Euclidean and Lorentzian cases.

There is a superficial similarity of our discussions about the monodromy of OPE blocks with other works that have considered monodromies. Some papers, such as \cite{Banerjee2016,Anous2016,Anous2017}, focus on correlators with large numbers of light operators in the background of two heavy insertions. Monodromy is used to relate the possible OPE channels of the overall correlator. Other papers, such as \cite{CarneirodaCunha2016,Maloney2017}, use monodromy as a way to pick out different channels of four point functions by switching heavy OPE exchanges with lighter ones. There are two main differences in what we have discussed. First, we are considering a single OPE block, not the full OPE, so the exchanged operators are fixed. All the works mentioned above involve multiple operators, which can fuse in different channels. In contrast the OPE block is a single operator; there is no notion of different fusion channels. Second, we implement sums to conform to the discrete gauge symmetry that is present on the base but not on the cover, which differs from the above works. 

It is also important to highlight a possible connection to entwinement \cite{Balasubramanian2015,Balasubramanian2016,Balasubramanian2019}. Entwinement has been proposed as the CFT dual to the length of non-minimal boundary anchored geodesics present in non-pure AdS spacetimes. Unlike the entanglement entropy of a boundary subregion, which is a measure of correlations among spatially organized degrees of freedom, these works suggest that entwinement measures correlations among internal, discretely gauged degrees of freedom. It seems likely that the entwinement/non-minimal geodesic length duality is closely related to the OPE block/geodesic integrated field duality, and it would be interesting to understand the deeper connections between them.

\section*{Acknowledgements}
We would like to thank Kanato Goto and A. Liam Fitzpatrick for useful discussions. This work is supported by a Discovery Grant from the Natural Sciences and Engineering Research Council (NSERC) of Canada. The work of JCC is also supported by a Vanier Canada Graduate Scholarship.

\bibliographystyle{JHEP}
\bibliography{CresswellJardinePeet-monodromyOPEblocks}

\end{document}